\documentclass[11pt,twoside]{article}

\usepackage{asp2014}
\usepackage{natbib}
\usepackage{listings}

\aspSuppressVolSlug
\resetcounters

\begin{document}

\title{The eOSSR library}

\author{Thomas~Vuillaume$^1$, Enrique~Garcia$^{1,2}$, Christian Tacke$^3$, and Tamas Gal$^4$}
\affil{$^1$Univ. Savoie Mont-Blanc, LAPP, CNRS, Annecy, France; \email{thomas.vuillaume@lapp.in2p3.fr}}
\affil{$^2$IT Department, CERN - 1211 Geneva 23 - Switzerland}
\affil{$^3$GSI Helmholtz Centre for Heavy Ion Research GmbH, Darmstadt, Germany}
\affil{$^4$IT Erlangen Centre for Astroparticle Physics, Erlangen, Germany}

\paperauthor{Vuillaume~T.}{thomas.vuillaume@lapp.in2p3.fr}{https://orcid.org/0000-0002-5686-2078}{Univ. Savoie Mont-Blanc, LAPP, CNRS}{}{Annecy}{}{74370}{France}
\paperauthor{Garcia~E.}{}{0000-0003-2224-4594}{CERN}{IT Department}{Geneva}{}{1211}{Switzerland}
\paperauthor{Tacke~C.}{}{0000-0002-5321-8404}{ GSI Helmholtz Centre for Heavy Ion Research GmbH}{}{Darmstadt}{}{}{Germany}
\paperauthor{Gal~T.}{}{0000-0001-7821-8673}{Erlangen Centre for Astroparticle Physics}{}{Erlangen}{}{}{Germany}



\begin{abstract}

The astronomy, astroparticle and particle physics communities are brought together through the ESCAPE (European Science Cluster of Astronomy and Particle Physics ESFRI research infrastructures) project to create a cluster focused on common issues in data-driven research. Among the ESCAPE work packages, the OSSR (ESCAPE Open-source Scientific Software and Service Repository) is a curated, long-term, open-access repository that makes it possible for scientists to exchange software and services and promote open science. It has been developed on top of a Zenodo community, connected to other services.
A Python library, the eOSSR, has been developed to take care of the interactivity between Zenodo, services and OSSR users, allowing an automated handling of the OSSR records. In this work, we present the eOSSR, its main functionalities and how it's been used in the ESCAPE context to ease the publication of scientific software, analysis, and datasets by researchers. 

\end{abstract}




\section{The ESCAPE OSSR}

The aim of the ESCAPE OSSR (Open-source Scientific Software and Service Repository) is to provide the tools necessary for the communities to share their science products in a harmonized way respecting the FAIR principles, promoting open science and maximizing cross-fertilization by software re-use and co-development. One of the key components to achieve this goal is the software and service repository.  
For its concept implementation, the ESCAPE repository is using Zenodo web service through the curated escape2020 community integrated with several tools to enable a complete software life-cycle. The ESCAPE Zenodo community welcomes entries that support the software and service projects in the OSSR such as user-support documentation, tutorials, presentations, and training activities. It also encourages the archival of documents and material that disseminate and support the goals of ESCAPE.

\section{The eOSSR library}

We developed a Python library called eOSSR \citep{eossr_zenodo} in order to allow an automated integration of Zenodo with other tools and platforms forming the OSSR as well as providing an integrated environment to external users.
The library is open-source and has been published in the OSSR itself. 
The documentation and running examples can be found online\footnote{\url{https://escape2020.pages.in2p3.fr/wp3/eossr/}}.

\begin{figure}
    \centering
    \includegraphics[width=\textwidth]{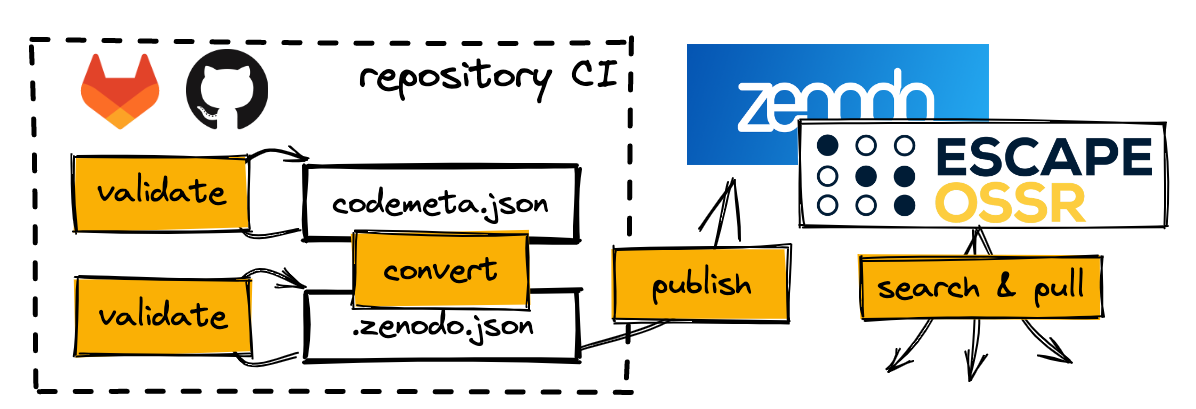}
    \caption{A typical software publication workflow in the OSSR using the eOSSR library}
    \label{fig:worflow}
\end{figure}

\subsection{Metadata}

Metadata: The OSSR has chosen codemeta.json as schema and format for its software metadata. The definition of this schema has been integrated within the eOSSR, thus allowing: 
An automated verification of the metadata.
A converter between Zenodo metadata schema and the OSSR metadata

\subsection{API}

For other tool to communicate with the OSSR, an API was necessary. The eOSSR takes advantage of Zenodo's API to propose a set of high-level functionalities through Python functions or command-lines, such as:
Requesting software in the OSSR via wide  or specific searches, using plain text or recognized metadata such as keywords or file type.

The \texttt{Record} object is at the core of the eOSSR Zenodo API. This class handles individual Zenodo records, containing all their metadata and useful methods to manipulate them. The easiest way to retrieve a Zenodo record is from it's \texttt{record\_id}:
\begin{lstlisting}
from eossr.api.zenodo import get_record
record_id = 6826881
record = get_record(record_id)
\end{lstlisting}

\noindent Here a some of the most useful methods associated to the \texttt{record} object:
\begin{itemize}
    \item \texttt{record.print\_info()}: prints general information (title, version, description, URL...) about the record to a stream
    \item \texttt{record.metadata}: all records metadata
    \item \texttt{record.get\_associated\_versions()}: retrieve all associated versions of that record published in Zenodo
    \item \texttt{record.get\_codemeta()}: directly reads the content of the \texttt{codemeta.json} file at the root of a record
    \texttt{record.write\_zenodo()}: writes the record metadata to a \texttt{.zenodo.json} file
\end{itemize}

\noindent To communicate with Zenodo and in particular with an user restricted access, one can use \texttt{zen = eossr.api.zenodo.ZenodoAPI(access\_token)} with its private access TOKEN provided by Zenodo\footnote{https://zenodo.org/account/settings/applications/tokens/new/}. It will then allow to:
\begin{itemize}
    \item \texttt{zen.get\_user\_records()}: returns all the \texttt{record}s owned by the user
    \item \texttt{zen.get\_community\_pending\_requests(community)}: returns a list of\\
    \texttt{record}s that have been requested to be added to a community.
    \item \texttt{zen.accept\_pending\_request(community, record\_id)}: accept a pending request into a community. The community must be owned by the token owner.
    \item \texttt{zen.create\_new\_entry()}: creates a new deposit Zenodo
    \item \texttt{zen.erase\_entry(entry\_id)}: erases a deposit that has not been published yet
    \item \texttt{zen.publish\_entry(entry\_id)}: publishes a deposit
    \item \texttt{zen.update\_record\_metadata(record\_id, metadata)}: updates a deposit metadata
    \item \texttt{zen.upload\_dir\_content(directory)}: packages the project root directory as a zip archive and upload it to Zenodo. If a record\_id is passed, a new version of that record is created. Otherwise, a new record is created.
\end{itemize}

\noindent The API also provides several general purpose functions to query Zenodo based on a text search:
\begin{itemize}
    \item \texttt{eossr.api.zenodo.zenodo.search\_records()}: general search function for records
    \item \texttt{eossr.api.zenodo.zenodo.search\_communities()}: general search function for communities
    \item \texttt{eossr.api.zenodo.zenodo.search\_funders()}: general search function for known funders
    \item \texttt{eossr.api.zenodo.zenodo.search\_grants()}: general search function for known grants
    \item \texttt{eossr.api.zenodo.zenodo.search\_licenses()}: general search function for known licenses
\end{itemize}

\noindent Specific OSSR functions are then built on this:
\begin{itemize}
    \item \texttt{eossr.api.ossr.get\_ossr\_pending\_requests()}
    \item \texttt{eossr.api.ossr.get\_ossr\_records()}
\end{itemize}

\subsection{Command line interface}

A set of command lines are also provided to ease the use of the eOSSR:

\paragraph{\texttt{eossr-codemeta2zenodo codemeta.json}}: Converts a metadata descriptive files from the the CodeMeta to the Zenodo schema. Creates a .zenodo.json file from a codemeta.json file.

\paragraph{\texttt{eossr-upload-repository --token TOKEN --input-dir DIRECTORY}}: Uploads a directory to the OSSR as record. The directory must include a valid zenodo or codemeta file to be used as metadata source for the upload. If not record\_id is passed, a new record is created. Otherwise, a new version of the existing record is created.

\paragraph{\texttt{eossr-metadata-validator codemeta.json}}: Validate a codemeta file. Raises warnings for recommended changes and errors for unvalid entries.

\paragraph{\texttt{eossr-check-connection-zenodo --token TOKEN --project\_dir DIRECTORY}}: Test the connection to zenodo and all the stages of a new upload. This is particularly useful in a set of CI tests to test that the upload to Zenodo will be done without issues when making a new release of the software.

\subsection{Code snippets for continuous integrations}

As one of the main usage of the eOSSR is in continuous integrations to insure automated upload to Zenodo (e.g. when making a software release), we also provide snippets in the documentation than you can use directly in your CI, notably in GitLab. This fills the missing GitLab-Zenodo\footnote{\url{https://escape2020.pages.in2p3.fr/wp3/eossr/gitlab_to_zenodo.html}} integration.

\acknowledgements The ASP would like to thank the dedicated researchers who are publishing with the ASP. ESCAPE - The European Science Cluster of Astronomy \& Particle Physics ESFRI Research Infrastructures has received funding from the European Union's Horizon 2020 research and innovation programme under Grant Agreement no. 824064.

\bibliography{eossr_bib}  


\end{document}